\documentclass[aps,twocolumn,showpacs]{revtex4}
\usepackage{amssymb}
\usepackage[dvips]{graphicx}
\usepackage{epsfig}

\begin{document}


\title{On the origin of the A$_{1g}$ and B$_{1g}$ electronic Raman scattering peaks
in the superconducting state of YBa$_{2}$Cu$_{3}$O$_{7-\delta}$}
\author{H. Martinho,$^{1}$ A. A. Martin,$^{2}$ C. Rettori,$^{1}$ C. T. Lin,$^{3}$ and
C. Bernhard,$^{3}$}

 \affiliation{$^{1}$Instituto de F\'{\i}sica ''Gleb Wataghin'', UNICAMP, 13083-970, Campinas, SP,
Brazil}
 \affiliation{$^{2}$Instituto de Pesquisa e Desenvolvimento - UNIVAP, 12 244-050, S\~{a}o Jos\'{e} dos Campos, SP,
Brazil}
\affiliation{$^{3}$Max-Planck-Institut f\"{u}r Festk\"{o}rperforschung, Heisenbergstrasse 1, D-70 569 Stuttgart,
Germany.\\}

\begin{abstract}
The electronic Raman scattering has been investigated in optimally oxygen
doped YBa$_{2}$Cu$_{3}$O$_{7-\delta}$ single crystals as well as in crystals
with non-magnetic, Zn, and magnetic, Ni, impurities. We found that the
intensity of the A$_{1g}$ peak is impurity independent and their energy to
$T_{c}$ ratio is almost constant ($2\Delta/k_{B}T_{c}\sim5$). Moreover, the
signal at the B$_{1g}$ channel is completely smeared out when non-magnetic Zn
impurities are present. These results are qualitatively interpreted in terms
of the Zeyher and Greco's theory that relates the electronic Raman scattering
in the A$_{1g}$ and B$_{1g}$ channels to \textit{d}-CDW and superconducting
order parameters fluctuations, respectively.

\end{abstract}


\maketitle

\draft

Far a long time, the phenomenon of high-$T_{c}$ superconductivity in the
cuprates has been considered as related to an unconventional pairing
state.\cite{Moriya} It is widely accepted that the unconventional mechanism in
these materials is closely related to their normal state properties, e.g., the
non-Fermi-liquid behavior,\cite{Chubukov} in spite of the lack of consensus
about the correct description of this state. The presence of several kinds of
fluctuations, such as spin, flux phases, stripes or charge density waves (CDW)
is an additional complication\cite{fluctuations} and many models considering
one or more of these fluctuations have been proposed. However, there is a
large amount of experimental data that these models cannot explain (see, e.g.,
sections 5.4 and 6.4 of Refs.\cite{Moriya} and \cite{Chubukov}, respectively).
Among these data we mention the absence of a convincent explanation for the
electronic Raman scattering (ERS) of many cuprates in the superconducting state.

The redistribution of the ERS in the superconducting state has been used to
study the gap order parameter in many superconductors, both conventional or
unconventional, such as Nb$_{3}$Sn,\cite{Nb} cuprates,\cite{cuprates}
borocarbides\cite{borocarbides} and recently, MgB$_{2}$.\cite{MgB2} In
general, the ERS of the cuprates presents two characteristic peaks in the
B$_{1g}$ and A$_{1g}+$B$_{2g}$ channels  in the superconducting phase that
disappear above $T_{c}$. In the YBa$_{2}$Cu$_{3}$O$_{7-\delta}$ (Y123) system
the ERS of optimally doped crystals has shown the A$_{1g}$ peak stronger than
the B$_{1g}$ counterpart. Also, the maximum intensity of the A$_{1g}$ peak
lies at a lower energy than that of the B$_{1g}$ one.\cite{AAM} In the
overdoped state, the B$_{1g}$ peak shifts down converging approximately to the
same position of the A$_{1g}$ peak\cite{AAM} while in the strongly underdoped
regime neither A$_{1g}$ nor B$_{1g}$ peaks are observed.

The appearance of these peaks only below $T_{c}$, leads one to believe that
they are related to the superconductivity, probably being pair-breaking peaks.
Some publications\cite{DV} have already shown that, for a gap with nodes and
$d_{x^{2}-y^{2}}$ order parameter at low energies, the ERS efficiency obeys
the $\sim\omega^{3}$ and $\sim\omega$ power-laws in the B$_{1g}$ and A$_{1g}%
+$B$_{2g}$ channels, respectively. Besides, these works have predicted that
the A$_{1g}$ peak should be weaker than the B$_{1g}$ component with their
maxima appearing at the same energy $2\Delta$. In spite of the good agreement
between the theory and the experimentally observed power laws, the relative
position and intensity of the ERS peaks are in disagreement with the
theoretically expected results for the pair-breaking ERS response. Hence, the
origin of the A$_{1g}$ and B$_{1g}$ peaks remains unclear.

\begin{figure}[th]
\includegraphics[height=8.6cm]{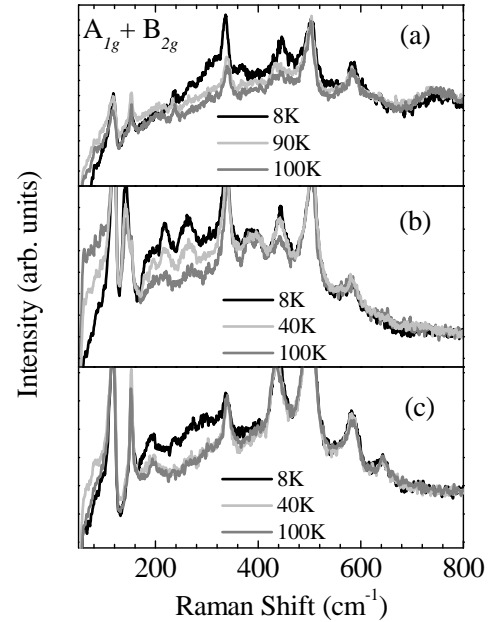}\caption{Temperature dependence of the
Raman spectra in the A$_{1g}$+B$_{2g}$ channel showing the redistribution of
the ERS below $T_{c}$ for (a) Y123 with T$_{c}=91$ K, (b) Y123:Ni with
T$_{c}=76$ K, and (c) Y123:Zn with T$_{c}=72$ K single crystals.}%
\label{fig1}%
\end{figure}

In this letter we make an attempt to clarify this question by means of ERS
measurements in Y123 single crystals doped with a small amount, 5\%, of
either, magnetic Ni$^{2+}$ (Y123:Ni) or non-magnetic Zn$^{2+}$ (Y123:Zn)
impurities in the copper planes. It is known that substituting Cu by Zn or Ni
in Y123 preserves the oxygen doping level and modify only slightly the crystal
structure. Also, the non-magnetic Zn$^{2+}$ impurities, contrary to the
magnetic Ni$^{2+}$ ones, restore significant spin fluctuations in the normal
state.\cite{Si,Fong} Since the presence of impurities in the CuO$_{2}$ planes
are known to induce a pair-breaking effect,\cite{Si,Fong} the investigation of
the change in the electronic properties by introducing magnetic and
non-magnetic impurities can give an important insight into the open questions
related to the physics of the cuprates.

Single crystals of Y123, Y123:Ni and Y123:Zn were prepared as previously
described.\cite{Lin} After a thermal treatment, the transition temperatures
were measured by \textit{dc}-magnetization and were found to be $91$, $76$ and
$72$ K for the Y123, Y123:Ni and Y123:Zn samples, respectively. In terms of
the oxygen level, all samples are in the optimally-doped state. The Raman
measurements were carried out using a triple spectrometer equipped with a
LN$_{2}$ CCD detector. The $514.5$ nm line of an Ar$^{+}$ ion laser was used
as an excitation source. The laser power at the sample was kept below $8$ mW
on a spot diameter of about $50$ $\mu$m. The samples were cooled in an
exchange He gas variable temperature cryostat, and measured in a
near-backscattering configuration on the $ab-$plane. For the tetragonal
symmetry $D_{4h}$, the choice of the $x^{\prime},x^{\prime}$ geometry probes a
combination of the A$_{1g}$ and $B_{2g}$ symmetries, while choosing the
$x^{\prime},y^{\prime}$ geometry couples to excitations of B$_{1g}$ symmetry.
$x^{\prime}$ ($y^{\prime}$) denote axes rotated by $45^{o}$ from the
crystallographic $x(y)$ axes.

\begin{figure}[t]
\includegraphics[height=8.6cm]{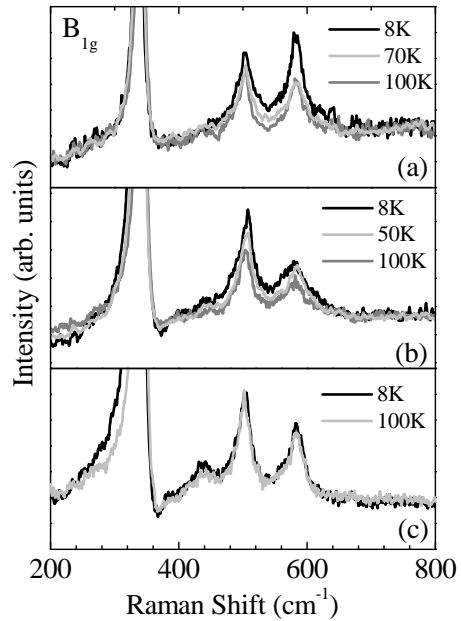}\caption{Temperature dependence of the
polarized Raman spectra in $B_{1g}$ channel for (a) Y123, (b) Y123:Ni and (c)
Y123:Zn.}%
\label{fig2}%
\end{figure}

In Fig.1 we present the Raman spectra in the A$_{1g}$+B$_{2g}$ channel at
different temperatures for the Y123, Y123:Ni and Y123:Zn samples, corrected by
the thermal Bose-Einstein factor. In Fig. 1(a) the spectrum for Y123 at $100$
K displays a flat background and just below T$_{c}\sim91$ K starts the
rearrangement of the electronic background, resulting in a broad peak in the
spectral range between $200$ and $400$ cm$^{-1}$. The same behavior is also
found in Fig.1 (b) for Y123:Ni. However, in this case the rearrangement of the
ERS starts only below $50$ K, producing a broad peak located in the same
spectral range as far the pure sample. For Y123:Zn, Fig.1(c), the gain of
spectral weight in the superconducting state is also present in the A$_{1g}%
$+B$_{2g}$ channel, although it starts to appears at $40$ K, below T$_{c}%
\sim72$ K, but is located in the same spectral range as the other two samples.

In the B$_{1g}$ channel, Fig.2, the rearrangement of the ERS is also displayed
for Y123 and Y123:Ni. For Y123, Fig.2(a), it starts below $70$ K and it
appears in the $450$ to $650$ cm$^{-1}$ spectral range. For Y123:Ni, Fig.2(b),
the broad peak first appears below $60$ K and it is also located in the $450$
to $650$ cm$^{-1}$ spectral range. Surprisingly, for Y123:Zn, Fig. 2(c), the
rearrangement of the ERS is absent below $T_{c}$. The only observed effect of
lowering the temperature is the broadening of the B$_{1g}$ phonon at $330$
cm$^{-1}$.

\begin{figure}[t]
\includegraphics[height=8.6cm]{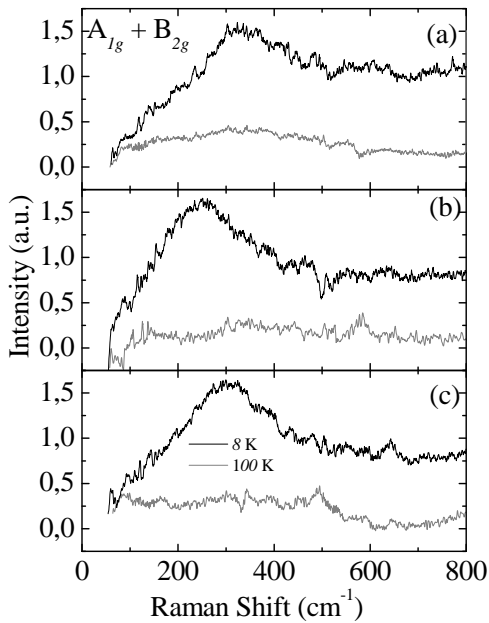}\caption[c3]{Electronic Raman response
at $8$ K (black lines) and $100$ K (gray lines) for (a)Y123, (b) Y123:Ni, and
(c)Y123:Zn single crystals in A$_{1g}+$B$_{2g}$ channel.}%
\label{fig3}%
\end{figure}

\begin{figure}[t]
\includegraphics[height=8.6cm]{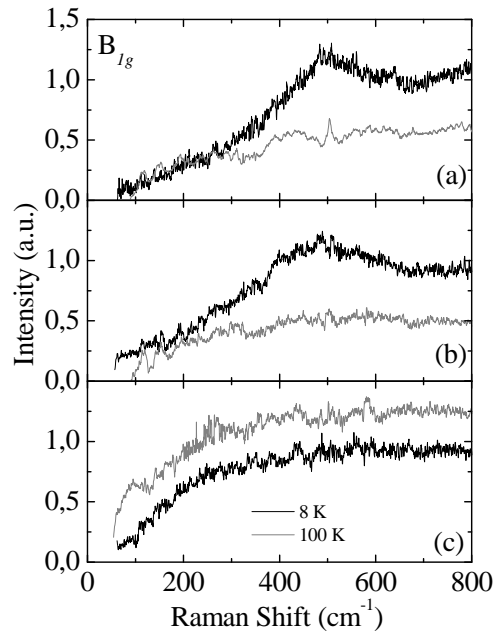}\caption[c4]{Electronic Raman response
at $8$ K (black lines) and $100$ K (gray lines) for (a) Y123, (b) Y123:Ni, and
(c) Y123:Zn in the B$_{1g}$ channel.}%
\label{fig4}%
\end{figure}

In order to determine the energy of the broad peaks which appear in the
superconducting phase in the A$_{1g}+$B$_{2g}$ and B$_{1g}$ channels, the pure
ERS response function has been obtained by subtracting the contribution of the
phonons fitted to Lorentzian or Fano-profiles to the Bose-Einstein corrected
raw data.

In Figure 3 we present the pure ERS response for the A$_{1g}+$B$_{2g}$ channel
at $8$ and $100$ K. At $8$ K we found the A$_{1g}$ response peaks around
$320$, $250$ and $300$ cm$^{-1}$ for Y123 (3a), Y123:Ni (3b) and Y123:Zn (3c),
respectively. We notice that the A$_{1g}$ energy to $T_{c}$ ratio $\hbar
\omega_{A_{1g}}/k_{B}T_{c}$ is $\sim5.0$ for Y123, $\sim4.7$ for Y123:Ni, and
$\sim5.8$ for Y123:Zn. As commented above, at $100$ K, the superconducting
rearrangement of the ERS is absent and the Raman spectra present no pronounced peak.

Figure 4 shows the pure ERS for the B$_{1g}$ channel at $8$ and $100$ K. The
dramatic difference between the response of Y123:Zn compared with that of the
other crystals is evident. The peak is absent in Y123:Zn but it appears
centered around $480$ cm$^{-1}$ in Y123 and Y123:Ni. The ratio $\hbar
\omega_{B_{1g}}/k_{B}T_{c}$ is $\sim7.7$ for Y123 and $\sim9.2$ for Y123:Ni.
Moreover, the ERS spectrum in Y123:Zn is almost the same at $8$ and $100$ K
being also very similar to those in Y123 and Y123:Ni at $100$ K, except for a
small difference in the absolute value of the intensities.

Observation of Fig. 3 and 4 indicates that the Zn substitution affects the
A$_{1g}+$B$_{2g}$ and B$_{1g}$ channels in a quite different way. While in the
A$_{1g}+$B$_{2g}$ channel the intensities of the peaks are almost unaffected
by the impurities, in the B$_{1g}$ channel the peak is smeared out in the
Zn-doped sample. The comparison between the B$_{1g}$ channel signals at $8$
and $100$ K in Y123:Zn indicates that the ERS in the superconducting and
normal states are nearly the same. This unusual effect of Zn substitution on
the B$_{1g}$ Raman response is surprising and has not been predicted by any
theoretical model nor been observed by other systematic experimental work.
Another relevant result is the almost constant energy to $T_{c}$ ratio
$\sim5-6$ for the A$_{1g}$ peak for all samples. This value is in agreement
with previous values obtained for the A$_{1g}$ peak\cite{gap,Gallais} and for
the superconducting gap measured by electron tunneling spectroscopy in
Y123.\cite{tunneling}

Recently, Venturini et al\cite{Vt} and Zeyher and Greco \cite{Greco} have
discussed theoretically the ERS in the superconducting phase of high-$T_{c}$ cuprates.

The work of Venturini et al.\cite{Vt} presents a model suggesting that the
observed peak in the A$_{1g}$ + B$_{2g}$ channel is produced by collective
spin fluctuations, being a two-magnon Raman peak. In the same model, the
B$_{1g}$ peak is related to pair breaking. They were able to fit the
experimental Raman spectrum of Bi2212 to their model and obtained the correct
relative position of the A$_{1g}$ and B$_{1g}$ peaks.\cite{Vt} Moreover,
Gallais et al.\cite{Gallais} have shown that the A$_{1g}$ peak tracks the
magnetic resonance peak observed by inelastic neutron scattering\cite{Si} at
40 $meV$ in Ni-substituted YBa$_{2}$Cu$_{3}$O$_{6.95}$. These authors have
interpreted this fact as an evidence for the magnetic origin of the A$_{1g}$ peak.

However, our results cannot be interpreted in these terms. The main reason is
the strong Zn-impurity dependency observed in the B$_{1g}$ spectra. In the
Venturini's framework the B$_{1g}$ peak is related to pair-breaking process
with their maximum at $2\Delta$ energy. Thus, the complete smearing out of the
B$_{1g}$ peak observed in our experiments (see Fig. 4 c) would imply that
$2\Delta\rightarrow0$ for the Zn-substituted crystal. However, it is known
that $T_{c}$ does not go to zero in this case and, discharging any anomalous
behavior of the $2\Delta/k_{B}T_{c}$ ratio, $2\Delta\nrightarrow0$.
Futhermore, the $\hbar\omega_{A_{1g}}/k_{B}T_{c}$ ratio for the A$_{1g}$ peak
presents better agreement to the gap energy measured by others groups (see,
e.g., ref. \cite{tunneling}) than the B$_{1g}$ one, indicating that the
A$_{1g}$ ERS peak is related to pair-breaking.

Another possible theoretical comparison could be made with the Zeyher and
Greco's\cite{Greco} theory. These authors used the superconducting model by
Cappelluti and Zeyher\cite{t-J} in order to understand the ERS of cuprates.
The Cappelluti and Zeyher model proposed that the superconductivity in
cuprates is originated by the competition between the superconducting and the
\textit{d}-CDW (also called orbital antiferromagnetism) order parameters.
Zeyher and Greco\cite{Greco} suggested that the A$_{1g}$ and B$_{1g}$ peaks
are caused by amplitude fluctuations of the superconducting and \textit{d}-CDW
order parameters, respectively. The \textit{d}-CDW phase corresponds to a flux
phase where current flows around each CuO$_{2}$ square alternatively clockwise
and counterclockwise\cite{d-CDW} giving rise to orbital antiferromagnetism
where the only interaction present in the order parameter is the Heisenberg
exchange coupling between the Cu$^{2+}$ ions.\cite{Greco}

As mentioned above, our experimental data indicates that the A$_{1g}$ peak is
related to pair-breaking. In this sense, our experimental ERS results give
support to the Zeyher and Greco's theory. Nevertheless, the impurity effect on
the \textit{d}-CDW B$_{1g}$ peak is not well theoretically understood at the
moment. Cappelluti and Zeyher\cite{impurities:t-J} have shown that the Zn
impurities substitution does not have appreciable influence on the oxygen
doping phase diagram of the cuprates. However, the effect on the B$_{1g}$ ERS
signal could be more subtle and a more detailed theoretical calculation could
be used.

Nonetheless, qualitatively we can elaborate about the impurity substitution
effect on B$_{1g}$ signal as follows. It is known that the main interaction
originating the \textit{d}-CDW is the Heisenberg coupling between the
Cu$^{2+}$ spins. Thus, it is expected that the Cu$^{2+}$ ($S=1/2$)
substitution by Zn$^{2+}$ ($S=0$) impurity, would break down the long range
coherence of the orbital antiferromagnetism, explaining the disappearence of
the \textit{d}-CDW excitation in the B$_{1g}$ channel. In fact, this is
consistent with the larger depletion of $T_{c}$ just produced by the presence
of  Zn$^{2+}$ impurities.\cite{Si,Fong} Besides, the Ni$^{2+}$ ($S=1$)
impurities could be less effective in breaking down the orbital
antiferromagnetism notwithstanding having also effect in decreasing $T_{c}$.
Moreover, as shown by Gupta and Gupta\cite{Gupta} the charge density
redistribution due to the Ni$^{2+}$ ions is localized while the Zn$^{2+}$
perform an extended perturbation. On this basis, one would expect that the
Zn$^{2+}$ ions would be more effective on reducing the \textit{d}-CDW excitations.

In conclusion, our results show that the A$_{1g}$ ERS peak present a constant
gap to $T_{c}$ ratio $\sim5-6$  independent of the presence of magnetic or
non-magnetic impurities. Also, the B$_{1g}$ ERS peak is smeared out in the
Y123 when the Cu$^{2+}$ is substituted by small amount of non-magnetic Zn
impurities whereas the A$_{1g}+$B$_{2g}$ spectra remain insensitive to the
kind of impurity. These results could be qualitatively interpreted in terms of
the of Zeyher and Greco's\cite{Greco} theory that relates the ERS in the
A$_{1g}$ and B$_{1g}$ channels to \textit{d}-CDW and superconducting order
parameters fluctuations, respectively.

\section{Acknowledgments}

This work was supported by the Brazilian Agencies CNPq and FAPESP. We would
like to thank A. Greco, P.G. Pagliuso and R. L. Doretto for fruitful discussions.

\end{document}